\input{aipcheck}

\documentclass[
    ,final            
  ]
  {aipproc}

\layoutstyle{6x9}

\begin{document}

\title{Small-scale hero: massive-star enrichment in the Hercules dwarf spheroidal}

\classification{98.52.Wz, 98.62.Bj}
\keywords      {Stars: abundances ---  stars: Population II  --- nuclear reactions, nucleosynthesis, abundances --- galaxies: evolution --- galaxies: dwarf --- galaxies: individual: Hercules}

\author{Andreas Koch}{
  address={Landessternwarte, Zentrum f\"ur Astronomie der Universit\"at Heidelberg, K\"onigstuhl 12, 69117 Heidelberg, Germany}
}

\author{Francesca Matteucci}{
  address={Dipartimento di Astronomia, Universit\'a di Trieste, via G. B. Tiepolo 11, 34143 Trieste, Italy}
}

\author{Sofia Feltzing}{
  address={Lund Observatory, Box 43, SE-22100 Lund, Sweden}
}

\begin{abstract}
Dwarf spheroidal galaxies are often conjectured to be the 
sites of the first stars. The best current contenders for 
finding the chemical imprints from the enrichment by those 
massive objects are the "ultrafaint dwarfs" (UFDs). 
Here we present evidence for remarkably low heavy element
abundances in the metal poor Hercules UFD.  
Combined with other peculiar abundance patterns this 
indicates that Hercules was likely only influenced by 
very few, massive explosive events -- thus bearing 
the traces of an early, localized chemical enrichment 
with only very little other contributions from other sources 
at later times. 
\end{abstract}

\maketitle


\section{Introduction}
Dwarf spheroidal (dSph) galaxies have always been characterized as very low-luminosity and presumably dark-matter dominated objects \cite{Mat98}. 
Over the past six or so years, numerous discoveries have dug deeply 
into  the realm of {\em ultra-}faint dSphs (UFDs), probing ever fainter magnitudes \cite{Koch09} and very high mass-to-light ratios \cite{Sim11}. 
The UFDs are predominantly old and metal poor systems,  nicely extending the metallicity-luminosity relation towards the faintest end \cite{Kir08}.  
While long evasive \cite{F04}, very metal poor stars below an [Fe/H] of $-3$ dex have now been uncovered in 
both the UFDs and the more luminous dSphs, with the current record holder lying at $-3.96$ dex \cite{Taf10}. 

At $M_V=-6.6$  mag, Hercules (Her) is one of the brighter UFDs. With a
high M/L ($\sim$300) its stellar mass is only
a few times 10$^4$ M$_{\odot}$  \cite{Mart08}. 
Narrow-band photometry \cite{A09} and medium- to high-resolution spectroscopy \cite{A09}; \cite{K08};  \cite{A11} have uncovered a broad range in ``metallicity'' in this galaxy with 
a full range in [Fe/H] from $-3.2$ to $-2$ dex. 

This presentation focuses on the results we obtained from the high-resolution spectroscopy of \cite{A11} (R=20000; 11 stars) and \cite{K08}
(R=20000; 2 stars). 
\subsection{[Ca/Fe]: hints on chemical evolution}
The Ca abundances reported by \cite{A11} trace the broad range found in [Fe/H], with [Ca/Fe] declining from the common halo plateau value around 0.4 dex down to subsolar values 
of $-0.4$  dex. These are summarized in Fig.~1 together with literature data for Galactic disk and halo stars \cite{V04} and luminous and ultrafaint UFDs (e.g., 
\cite{Koch09}). 
\begin{figure}[htb]
  \includegraphics[width=.7\hsize]{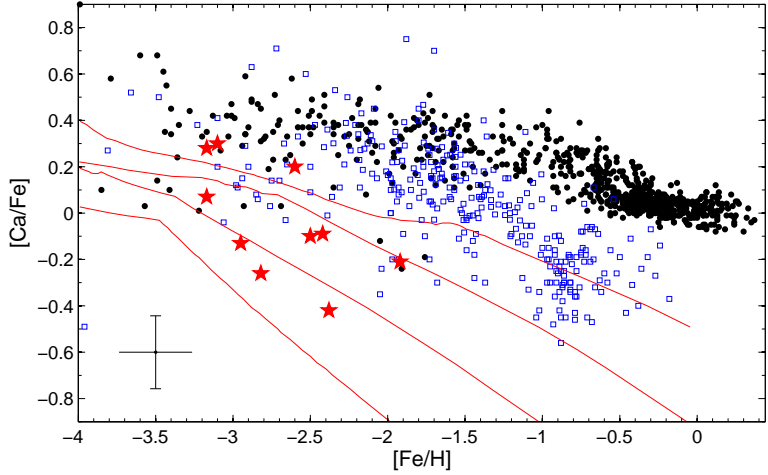}
  \caption{[Ca/Fe] ratio in MW disk and halo stars (black dots), dSph stars (blue squares) and our Her targets (red symbols). We also overplot models of chemical evolution \cite{LM04}, computed 
  for star forming efficiencies of (top to bottom) $\nu$=0.1, 0.01, 0.001, and 0.0001 Gyr$^{-1}$.}
\end{figure}

Fig.~1 also shows chemical evolution models specific to Her \cite{LM04}. While, clearly, more detailed calculations are required such as to, e.g., pin down the role of galactic winds, 
the models indicate that a low star forming efficiency of 0.001--0.01 Gyr$^{-1}$ can plausibly match the observed trend.
\subsection{[Mg/Ca]: the role of massive stars}
\cite{K08} reported on a very high ratio of the hydrostatic to explosive $\alpha$-elements (represented by Mg vs. Ca) in two of the more metal rich stars in Her (at [Fe/H]$\sim -2$ dex).  
This is a key prediction of nucleosynthetic models for type II Supernovae (SNe) that arise from massive progenitors. This is shown  in Fig.~2 exemplary for the metal-free SNe models of 
\cite{HW10}. 
\begin{figure}[htb]
  \includegraphics[width=.7\hsize]{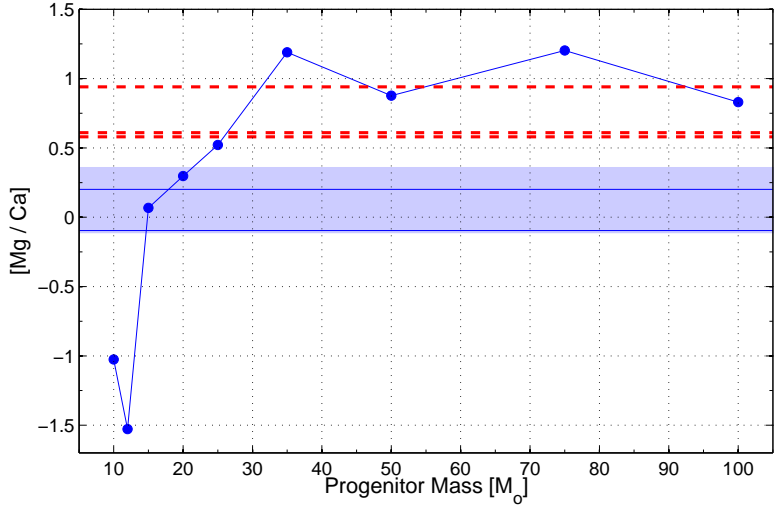}
  \caption{[Mg/Ca] ratios as observed in (dashed lines; top to bottom) two stars in Her and Dra 119. The blue points are nucleosynthetic yields vs. progenitor mass, adopting the models 
  of \cite{HW10}. The horizontal blue lines and shaded area indicate the 1$\sigma$-range in the Galactic halo and dSph stars, respectively.}
\end{figure}
The implied mass range to achieve the  high ratios in the Her stars (and Dra 119, which shows abundance patterns similar to the Her stars; \cite{F04}) 
is  of the order of 25--50 M$_{\odot}$. While we are thus dealing with an early enrichment by very short-lived, massive stars, 
this  mass range can still be considered ``ordinary massive'' (N. Yoshida; this proceedings) compared to early notions of very massive Population~III stars.

The [Mg/Ca] ratios in Her exceed the average values found  in Galactic halo stars by more than 2$\sigma$ 
and only a few peculiar halo stars show similar relative enhancements. Moreover, this is only seen in 3 stars in 2 other UFDs so far (Boo~I; \cite{Fel09}; Leo IV; \cite{Sim10}). 
Unfortunately, the spectral range  of \cite{A11} did not permit the measurement of Mg abundances.
\subsection{[Ba/H]: a puzzling deficiency}
The site of the $r$-process is still under debate, possibly occurring only in a limited range of SNe II masses  around 8--10 M$_{\odot}$ \cite{Q03}. 
The discovery of a strong depletion in essentially all heavy (Z$>$28) 
elements in the metal poor 
dSph star Dra 119 (\cite{F04}) suggest that more massive progenitors (as mandated by that star's light element patterns)
 in fact do not contribute significantly to the production of the $n$-capture elements. 
This is on par with the  deficiency in the heavy elements seen in the two stars of \cite{K08}. Furthermore, the FLAMES spectra of \cite{A11} allowed us to (only) place upper limits on the Ba-abundance in a further 11 red giants. These are shown in Fig.~3. 
\begin{figure}[htb]
  \includegraphics[width=.7\hsize]{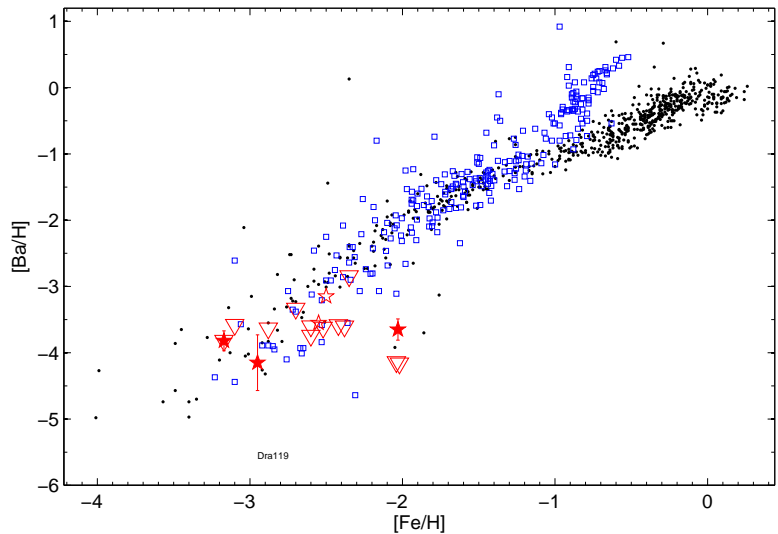}
  \caption{[Ba/H] abundances in the same environments as shown in Fig.~1. Red solid symbols reflect values based on equivalent widths 
  measurements, while open symbols refer to upper (3$\sigma$) detection limits in 
  the Her stars (triangles: this work; open stars: \cite{Fr12}). }
\end{figure}

As a result, Ba in Her is very low, at  [Ba/H]$<-3$ dex (Koch, Ad\'en, \& Feltzing 2012, in prep.). 
Moreover, there is no trend seen in Ba/H with [Fe/H] so that  whatever mechanism enriched this UFD did not produce any significant $n$-capture elements.
We also overplot the recent low-resolution measurements of \cite{Fr12} on Fig.~3. While they 
 ``did not find extreme abundances in our Hercules stars as the one found by \cite{K08}'',  
their Ba-limits are in fact well consistent with our values, albeit at slightly lower metallicities. 
\section{Discussion}
The key question is, how likely is the occurrence of those massive-star events in a low- (stellar) mass environment like Her?
\cite{K08} argued that, in Her, we are facing a highly incomplete sampling of the high-mass end of the galaxy's IMF in small-scale star formation events: 
in fact, our stochastic tests imply that  1--3 ``ordinary massive'' ($\sim35$ M$_{\odot}$) SNe II are sufficient to reproduce the observed patterns. 
Coupled with 
inhomogeneous mixing of the SN ejecta, Her's chemical enrichment was then governed by only a few massive stars, which boosted the initial [Ca/Fe] to the plateau value 
and produced the high Mg/Ca ratios. 
Those stars  would not synthesize any $n$-capture elements. Only  later, simmering star formation \cite{LM04} 
can then induce the broad, observed Fe spread. 
%
%
%
\vspace{-2ex}
\begin{theacknowledgments}
AK acknowledges the Deutsche Forschungsgemeinschaft for funding from  Emmy-Noether grant  Ko 4161/1. 
\end{theacknowledgments}



\vspace{-2ex}
\bibliographystyle{aipproc}   


\bibliography{sample}


\end{document}